\documentclass[12pt]{article}

\usepackage{amsmath}

\begin{document}

\begin{center}
{\large \textbf{Hidden Symmetry, Excitonic Transitions and Two-Dimensional
Kane's Exciton in the Quantum Well}}
\end{center}

\medskip

\begin{center}
\textbf{E.M. Kazaryan}$^{1}$\textbf{, L.S. Petrosyan}$^{1,2}$\textbf{, H.A.
Sarkisyan}$^{\ast 1,2}$

\emph{$^{1}$ Physicotechnical Department, Russian-Armenian State University,
123 Hovseph Emin str., Yerevan 375051, Republic of Armenia},\\[0pt]
\emph{$^{2 }$Department of Solid State Physics, Yerevan State University, 1
Alex Manoogian str., Yerevan 375025, Republic of Armenia},\\[0pt]
\emph{$^{\ast }$Corresponding author, e-mail: \underline {shayk@ysu.am},
Phone: +374 93 524718}
\end{center}

\medskip

\centerline{\bf Abstract}

In this article it is shown that, Sommerfeld's coefficients for excitonic
transitions in quantum wells are determined only with the principle quantum
number within the framework of two-dimensional Coulomb potential. This is a
consequence of hidden symmetry of two-dimensional Coulomb problem,
conditioned by the existence of two-dimensional analog of the Runge-Lentz
vector. For the narrow gap semiconductor quantum well with the non-parabolic
dispersion law of electron and hole in the two-band Kane model it is shown
that two-dimensional excitonic states are described in the frames of an
analog of Klein-Gordon equation with the two-dimensional Coulomb potential.
The non-stability of the ground state of the two-dimensional Kane's exciton
is show.

\medskip



\bigskip

\begin{center}
\textbf{1. Introduction}
\end{center}

Due to the existence of the size quantization effects in the semiconductor
nanostructures it became possible to realize such a low dimensional systems,
which initially had only model character. One of such is the two-dimensional
Coulomb system, which is obtained during the impurity and excitonic states
formation in quantum wells and superlattices [1-8]. Mathematically
two-dimensional Coulomb potential is defined by the following expression [9]:

\begin{equation}  \label{eq1}
V\left( {x,y} \right) = - \frac{Ze^2}{\sqrt {x^2 + y^2} },
\end{equation}

\noindent where $Z - $ is the charge number. Corresponding Schr\"odinger
equation in the Cartesian coordinates is the following

\begin{equation}  \label{eq2}
- \frac{\hbar ^2}{2m}\left( {\frac{\partial ^2}{\partial x^2} + \frac{%
\partial ^2}{\partial y^2}} \right)\Psi - \frac{Ze^2}{\sqrt {x^2 + y^2} }%
\Psi = E\Psi .
\end{equation}

It should be mentioned that two-dimensional hydrogen atom, being a system
with the hidden symmetry, can be investigated simultaneously in several
coordinate systems as: polar [10], parabolic [11] and elliptical [12]. In
the absence of fields, it is more appropriate to use polar coordinate system.

As it was mentioned, one of the causes for the formation of two-dimensional
Coulomb systems in semiconductor quantum wells is the formation of excitons
in them during the interband optical absorption [13]. Since the system under
is two-dimensional the exciton in it is also two-dimensional (provided
strong size quantization [14]) and the energy of electron-hole interaction
can be described in the frames of two-dimensional Coulomb potential (\ref%
{eq1}) and charge number $Z = 1$. Two-dimensional excitonic systems can be
realized also in quantum dots in the presence of strong vertically
quantization or magnetic field (see f. e. [15-17]). Thus, two-dimensional
excitonic states will also have the hidden symmetry, natural for the
two-dimensional Coulomb problem. It is clear that this fact should be
reflected on the character of the excitonic transitions in quantum well.

On the other hand it should be mentioned that not all semiconductor quantum
wells have the parabolic dispersion law of the charge carriers. There are
compounds in which, due to the existence of strong interaction of valence
band with the band of conductivity, non parabolic dispersion law for the
electrons and holes can exist [18]. For that case, theoretical description
of the two-dimensional excitons should be done with the consideration of the
kinetic operator, which deviates from standard quadratic. It is natural to
expect that the above mentioned case can bring to the disappearance of the
occasional degeneration of the two-dimensional Coulomb problem discussed.

In this article we study the influence of the hidden symmetry on the
Sommerfeld coefficient for excitonic transitions in semiconductor quantum
wells. We also investigate two-dimensional excitonic states in narrowband
semiconductor quantum wells with the Kane's dispersion low for charge
carriers.

\medskip

\begin{center}
\textbf{2. Two-dimensional excitonic states}
\end{center}

As was mentioned above we are studying the two-dimensional electron-hole
system with the interaction potential (\ref{eq1}). Corresponding Schr\"odinger
two-particle equation will have the following form.

\begin{equation}  \label{eq3}
- \left\{ {\frac{\hbar ^2}{2m_e }\Delta _e + \frac{\hbar ^2}{2m_h }\Delta _h
+ \frac{e^2}{\left| {\vec {\rho }_e - \vec {\rho }_h } \right|}}
\right\}\Psi = E\Psi ,
\end{equation}

\noindent where $m_{e\left( h \right)} - $is the effective mass of the
electron(hole) and $\frac{\hbar ^2}{2m_{e\left( h \right)} }\Delta _{e\left(
h \right)} - $ is the kinetic energy of the electron(hole) operator.
Introducing reduced mass

\begin{equation*}
\mu = \frac{m_e m_h }{m_e + m_h },
\end{equation*}

Relative coordinate

\begin{equation*}
\vec {\rho } = \vec {\rho }_e - \vec {\rho }_h ,
\end{equation*}

\noindent and applying standard procedure for the transition from two- to
one-particle problem we will obtain two-dimensional Schr\"odinger
equation

\begin{equation}  \label{eq4}
- \left\{ {\frac{\hbar ^2}{2\mu }\Delta + \frac{e^2}{\rho }} \right\}\Psi =
E\Psi .
\end{equation}

In polar coordinates equation (\ref{eq4}) will get to the following form [9]

\begin{equation}  \label{eq5}
- \frac{\hbar ^2}{2\mu }\left\{ {\frac{1}{\rho }\frac{\partial }{\partial
\rho }\left( {\rho \frac{\partial }{\partial \rho }} \right) + \frac{1}{\rho
^2}\frac{\partial ^2}{\partial \varphi ^2}} \right\}\Psi _{n,m} \left( {\rho
,\varphi } \right) - \frac{e^2}{\rho }\Psi _{n,m} \left( {\rho ,\varphi }
\right) = E\Psi _{n,m} \left( {\rho ,\varphi } \right).
\end{equation}

The well-known solution for this equation is following

\begin{equation}  \label{eq6}
\Psi _{n,m} \left( {\rho ,\varphi } \right) = \frac{e^{im\varphi }}{\sqrt {%
2\pi } }R_{n,m} \left( \rho \right),
\end{equation}

\noindent where

\begin{equation}  \label{eq7}
R_{n,m} \left( \rho \right) = \left[ {\frac{2\left( {n - \left| m \right|}
\right)!}{a_{ex}^2 \left( {n + \frac{1}{2}} \right)^3\left\{ {\left( {n +
\left| m \right|} \right)!} \right\}^3}} \right]^{\frac{1}{2}}\left( {\frac{%
\rho }{a_{ex} \lambda }} \right)^{\left| m \right|}e^{ - \left( {\frac{\rho
}{2a_{ex} \lambda }} \right)}L_{n + \left| m \right|}^{2\left| m \right|}
\left( {\frac{\rho }{a_{ex} \lambda }} \right),
\end{equation}

\noindent and where $a_{ex} = \frac{\hbar ^2}{\mu e^2} - $is the exciton
radius, $\lambda = \sqrt { - \frac{Ry^\ast }{4E}} $, $Ry^\ast - $ is the
effective Rydberg's energy, $m - $is the magnetic quantum number, $n =
n_\rho + \left| m \right| - $is the principle quantum number ($m = 0,\pm
1,\pm 2,...,\pm n)$, $n_\rho - $is the radial quantum number, $L_n^\alpha
\left( x \right) - $is the Laguerre's polynomial. Corresponding energetic
spectrum will be

\begin{equation}  \label{eq8}
E_n^{ex} = - \frac{\mu e^4}{2\hbar ^2\left( {n + \frac{1}{2}} \right)^2}.
\end{equation}

As it can be seen from (\ref{eq8}), two-dimensional excitonic states are
degenerated with the degeneracy order

\begin{equation}  \label{eq9}
g\left( n \right) = 2n + 1.
\end{equation}

Instead of expected $O\left( 2 \right)$ rotational symmetry (with respect to
$z$ axis) given by the matrix

\begin{equation}  \label{eq10}
R\left( \varphi \right) = \left( {{\begin{array}{*{20}c} {\cos \varphi }
\hfill & { - \sin \varphi } \hfill & 0 \hfill \\ {\sin \varphi } \hfill &
{\cos \varphi } \hfill & 0 \hfill \\ 0 \hfill & 0 \hfill & 1 \hfill \\
\end{array} }} \right),
\end{equation}

\noindent the considered system must possess more large symmetry. In the
paper [19] it is shown, that this degeneration is connected with the
existence of the two-dimensional analog of the Runge-Lentz vector in the
two-dimensional Coulomb problem. Components of that vector are

\begin{equation}  \label{eq11}
\begin{array}{l}
\hat {\gamma }_1 = \frac{1}{\sqrt { - 2E} }\left( {\frac{x}{\rho } - \hat {P}%
_y \hat {L}_z + i\frac{\hat {P}_x }{2}} \right), \\
\hat {\gamma }_2 = \frac{1}{\sqrt { - 2E} }\left( {\frac{y}{\rho } + \hat {L}%
_z \hat {P}_x - i\frac{\hat {P}_y }{2}} \right). \\
\end{array}%
\end{equation}

These operators along with plane rotation generator $\hat {\gamma }_3 =
\hat {L}_z = - i\hbar \frac{\partial }{\partial \varphi }$ Abelian group $%
O\left( 2 \right)$ define hidden group symmetry $SO\left( 3 \right)$
commutating with the Hamiltonian of the two-dimensional Coulomb problem and
providing corresponding degeneracy order. At the same time generators of
this group satisfy the commutation relation

\begin{equation}  \label{eq12}
\left[ {\hat {\gamma }_i \hat {\gamma }_j } \right] = i\varepsilon _{ijk}
\hat {\gamma }_k ,
\end{equation}
where $\varepsilon _{ijk}$ is the absolute anti-symmetric tensor.

\begin{center}
\textbf{3. Sommerfeld coefficients and excitonic transitions in quantum well}
\end{center}

During the investigation of the optical characteristics of the
semiconductors it was revealed that under the certain conditions of the
light absorption in semiconductors may be obtained such a conditions in
which electron transiting from valence band to the conductivity band can
create a bound state with the hole in the valence band, without reaching the
conductivity band [14]. In other words excitonic state occurs and
corresponding transitions are called excitonic. In the theory of
semiconductor optical properties it is shown that the intensity of the
excitonic transitions is characterized with the Sommerfeld coefficients
[14], which are equal to the values of the squared modulus of excitonic wave
function in $\rho = 0$

\begin{equation}  \label{eq13}
Z_1 \left( n \right) = \left| {\Psi _{n,m} \left( 0 \right)} \right|^2,
\end{equation}

\noindent if $\Psi _{n,m} \left( 0 \right) \ne 0$, and

\begin{equation}  \label{eq14}
Z_2 \left( n \right) = \left| {grad\Psi _{n,m} \left( 0 \right)} \right|^2,
\end{equation}

\noindent if $\Psi _{n,m} \left( 0 \right) = 0$. For the first case
transitions are called allowed, and for the second case -- forbidden.

Let's define the Sommerfeld coefficients for the excitonic transitions in
the quantum well, when exciton is described in the frames of above mentioned
two-dimensional Coloumbic problem. We should use the wave function (\ref{eq7}%
). From that it is obvious that allowed transitions take place only for the
states for which$m = 0$. Thus the corresponding Sommerfeld's coefficient
will have the following form [14]

\begin{equation}  \label{eq15}
Z_1 \left( n \right) = \left| {\Psi _{n,0} \left( 0 \right)} \right|^2 =
\frac{1}{\pi a_{ex}^2 \left( {n + \frac{1}{2}} \right)^3}.
\end{equation}

For the forbidden transitions case, first, we should write Laguerre's
polynomial in the waive function in the (\ref{eq7}) explicit form

\begin{equation}  \label{eq16}
L_{n + \left| m \right|}^{2\left| m \right|} \left( x \right) =
\sum\limits_{k = 0}^{n - \left| m \right|} {\frac{\left( {\ - 1} \right)^{k
+ 2\left| m \right|}\left\{ {\left( {n + \left| m \right|} \right)!}
\right\}^2x^k}{\left( {n - \left| m \right| - k} \right)!\left( {2\left| m
\right| + k} \right)!k!}} .
\end{equation}

Direct calculations show that $grad\Psi _{n,m} \left( 0 \right)$ is nonzero
when $m = 0$ and $m = \pm 1$. So for the Sommerfeld's coefficient we can take

\begin{equation}  \label{eq17}
Z_2 \left( n \right) = \sum\limits_{i = - 1}^1 {\left| {grad\Psi _{n,i}
\left( 0 \right)} \right|} ^2.
\end{equation}

Calculating the corresponding values for the gradients we can see that

\begin{equation}  \label{eq18}
\left| {grad\Psi _{n,0} \left( 0 \right)} \right|^2 = \frac{4\left( {n +
\frac{1}{2}} \right)^2}{\pi a_{ex}^4 \left( {n + \frac{1}{2}} \right)^5},
\end{equation}


\begin{equation}  \label{eq19}
\left| {grad\Psi _{n,\pm 1} \left( 0 \right)} \right|^2 = \frac{n\left( {n +
1} \right)}{\pi a_{ex}^4 \left( {n + \frac{1}{2}} \right)^5}.
\end{equation}

Taking into account (17-19) for the Sommerfeld's coefficient in the case of
forbidden transitions we get:

\begin{equation}  \label{eq20}
Z_2 \left( n \right) = \frac{3\left( {n + \frac{1}{2}} \right)^2 - \frac{1}{4%
}}{2\pi a_{ex}^4 \left( {n + \frac{1}{2}} \right)^5}.
\end{equation}

As it follows the intensity of the forbidden excitonic transitions in the
two-dimensional case depends only on the principle quantum number $n$. For
its fixed values, due to the existence of hidden symmetry of the discussed
problem, for the states with different $m$ and $n_\rho $, but with the same $%
n = n_\rho + \left| m \right|$ we get the same intensity for the excitonic
transitions $Z_2 \left( n \right)$. In Table 1 Sommerfeld's coefficients for
different forbidden transitions are brought.

\begin{table}[htbp]
\begin{tabular}{|p{96pt}|p{104pt}|p{109pt}|}
\hline
$n = 1$ & $n = 2$ & $n = 3$ \\[2pt] \hline
$n_\rho = 1,\,m = 0$ & $n_\rho = 2,\,m = 0$ & $n_\rho = 3,\,m = 0$ \\%
[2pt] \hline
$n_\rho = 0,\,m = 1$ & $n_\rho = 1,\,m = 1$ & $n_\rho = 2,\,m = 1$ \\%
[2pt] \hline
$n_\rho = 0,\,m = - 1$ & $n_\rho = 1,\,m = - 1$ & $n_\rho = 2,\,m = - 1$ \\%
[2pt] \hline
&  &  \\[-8pt]
$Z_2 \left( 1 \right) = \frac{1}{\pi a_{ex}^4 }\frac{104}{243}$ & $Z_2
\left( 2 \right) = \frac{1}{\pi a_{ex}^4 }\frac{296}{3125}$ & $Z_2 \left( 3
\right) = \frac{1}{\pi a_{ex}^4 }\frac{584}{16807}$ \\[4pt] \hline
\end{tabular}
\label{tab1}
\caption{Sommerfeld's coefficients for the forbidden two-dimensional
excitonic transitions for different values of main quantum number $n$.}
\end{table}

To sum up, we can conclude that for the strong quantization in quantum well,
when we can use two-dimensional exciton model, due to the hidden symmetry of
the two-dimensional Coulomb problem, Zommerfield's coefficients of the
exciton transitions are expressed only with principle quantum number. This
makes possible for the case of forbidden transitions to realize the same
intensity of the two-dimensional excitonic transitions for different states.
As soon as we assume the quasi two-dimensionality of the exciton and insert
the z coordinate into Hamiltonian the hidden symmetry of the problem
disappears and Sommerfeld's coefficients now depend also on $m$ and $n_\rho $%
.

\begin{center}
\textbf{4. Two-dimensional Klein-Gordon equation and Kane's exciton in
quantum well}
\end{center}

Along with the consideration of quasi two-dimensionality, also consideration
of non-parabolicity of dispersion law for the electron and hole in the
narrow band semiconductor quantum wells in two-dimensional Coulomb problem
can bring to the disappearance of the hidden symmetry. What is important,
that electron-whole interaction potential can be described within the frames
of two-dimensional potential (\ref{eq1}). In Kane's works it was shown that
consideration of interband interactions in semiconductor brings to the
deviation of the electron-whole dispersion law from the quadratic form [18].
In two-band approximation, when the effective masses of electron and whole
are equal, Kane's dispersion law becomes analogous to the relativistic one,
however, naturally, there is nothing relativistic, simple mathematical
coincidence. If we introduce the band interaction parameter $s(s \approx 10^8
$ sm/s), which is defined through interband dipole matrix element, the
corresponding dispersion law will take the following form [18]:

\begin{equation}  \label{eq21}
E = \sqrt {p^2s^2 + \mu ^2s^4} .
\end{equation}

Thus, within the frames of two-band Kane's approximation for defining the
excitonic states one does have to solve corresponding steady-state
Klein-Gordon equation with the Coulomb potential. It is clear, that for the
two-dimensional excitonic states in narrowband semiconductor quantum well
problem reduces to the investigation of the two-dimensional Klein-Gordon
equation with the Coulombic interaction term (\ref{eq1}).

In the polar coordinates Klein-Gordon equation for the Coulomb field takes
the following form [20]

\begin{equation}  \label{eq22}
\left[ {\frac{1}{\rho }\frac{\partial }{\partial \rho }\left( {\rho \frac{%
\partial }{\partial \rho }} \right) + \frac{1}{\rho ^2}\frac{\partial ^2}{%
\partial \varphi ^2}} \right]\Psi _{n_\rho ,m} + \left[ {\frac{2ZEe^2}{\hbar
^2c^2\rho } + \frac{Z^2e^4}{\hbar ^2c^2\rho ^2} - \frac{1}{\hbar ^2c^2}%
\left( {\mu ^2c^4 - E^2} \right)} \right]\Psi _{n_\rho ,m} = 0,
\end{equation}
where $\mu $ is the electron mass.

Representing electron's wave function as a compound of radial $R(\rho )$ and
angular $\Phi \left( \varphi \right)$ functions $\Psi _{n_\rho ,m} \left( {%
\rho ,\varphi } \right) = R_{n_\rho ,m} \left( \rho \right)\Phi _m \left(
\varphi \right)$ and using the variable separation method, from (\ref{eq1})
we obtain two equations which define $R(\rho )$ and $\Phi \left( \varphi
\right)$:
\begin{equation}  \label{eq23}
\rho ^2\frac{d^2R_{n_\rho ,m} }{d\rho ^2} + \rho \frac{dR_{n_\rho ,m} }{%
d\rho } + \left[ {\frac{2Ze^2E}{\hbar ^2c^2}\rho + \frac{Z^2e^4}{\hbar ^2c^2}
- \frac{\left( {\mu ^2c^4 - E^2} \right)}{\hbar ^2c^2}\rho ^2 - m^2} \right]%
R_{n_\rho ,m} = 0,
\end{equation}
and
\begin{equation}  \label{eq24}
\frac{d^2\Phi _m }{d\varphi ^2} + m^2\Phi _m = 0,
\end{equation}
where $m = 0;\,\pm 1;\,\pm 2;\,...$ is the magnetic quantum number, $n_\rho
- $is the radial quantum number.

From the (\ref{eq24}) for the normalized angular functions $\Phi _m \left(
\varphi \right)$ we get
\begin{equation}  \label{eq25}
\Phi _m \left( \varphi \right) = \frac{1}{\sqrt {2\pi } }e^{im\varphi }.
\end{equation}

For the radial equation (\ref{eq23}) first, let's introduce following
notations:
\begin{equation}  \label{eq26}
\varepsilon = \frac{1}{\hbar c}\left( {\mu ^2c^4 - E^2} \right)^{1 %
\mathord{\left/ {\vphantom {1 2}} \right. \kern-\nulldelimiterspace}
2},\quad \lambda = Z\alpha E\left( {\mu ^2c^4 - E^2} \right),\quad \alpha =
\frac{e^2}{\hbar c},
\end{equation}
and dimensionless variable $r = 2\varepsilon \rho $. Then for the $R\left( r
\right)$we come to the following equation
\begin{equation}  \label{eq27}
{R}^{\prime\prime}_{n_\rho ,m} + \frac{{R}^{\prime}_{n_\rho ,m} }{r} +
\left( {\frac{\lambda }{r} + \frac{Z^2\alpha ^2 - m^2}{r^2} - \frac{1}{4}}
\right)R_{n_\rho ,m} = 0.
\end{equation}

Below we find the solutions of equations (\ref{eq27}) in two limiting cases:
$r \to 0$ and $r \to \infty $.  Then the equation for $R\left( r \right)$ is
reduced to

\begin{equation}  \label{eq28}
{R}^{\prime}+ \frac{1}{r}{R}^{\prime}+ \frac{Z^2\alpha ^2 - m^2}{r^2}R = 0.
\end{equation}

We look for $R\left( r \right)$ in the form $r^S$. Substituting this into (%
\ref{eq28}) we obtain the following quadratic equation for $S$

\begin{equation}  \label{eq29}
S(S - 1) + S + Z^2\alpha ^2 - m^2 = 0.
\end{equation}

The two roots are:

\begin{equation*}
S_1 = \sqrt {m^2 - Z^2\alpha ^2} ,\quad S_2 = - \sqrt {m^2 - Z^2\alpha ^2} .
\end{equation*}

Only the first root guaranties the convergence of $R\left( r \right)$ at $r
\to 0$. Finally, we have:

\begin{equation}  \label{eq30}
R(r) = R_0 \left( r \right)\sim r^{\sqrt {m^2 - Z^2\alpha ^2} }.
\end{equation}

$r \to \infty .$In this case the equation (\ref{eq28}) take the form:

\begin{equation}  \label{eq31}
{R}^{\prime}- \frac{1}{4}R = 0.
\end{equation}

The solution obeying the standard conditions is:

\begin{equation}  \label{eq32}
R(r) = R_\infty (r)\sim e^{ - r \mathord{\left/ {\vphantom {r 2}} \right.
\kern-\nulldelimiterspace} 2}.
\end{equation}

In general case we apply for $R\left( r \right)$ the ansatz $R_\infty \cdot
R_0 \cdot u\left( r \right)$, where $u(r)$ is unknown function. So,

\begin{equation}  \label{eq33}
R\left( r \right) = r^S\exp \left( {{\ - r} \mathord{\left/ {\vphantom {{ -
r} 2}} \right. \kern-\nulldelimiterspace} 2} \right)u\left( r \right).
\end{equation}

Substituting (\ref{eq33}) in (\ref{eq27}) we get for $u(r)$ Kummer equation
[21]:

\begin{equation}  \label{eq34}
r{u}^{\prime}+ \left( {2S + 1 + r} \right){u}^{\prime}+ \left( {\lambda - S
- \frac{1}{2}} \right)u = 0.
\end{equation}

We will find the solution of (\ref{eq34}) in the form of power series:

\begin{equation}  \label{eq35}
u(r) = \sum\limits_{k = 0}^\infty {a_k r^k} .
\end{equation}

Substituting (\ref{eq35}) in (\ref{eq34}) and gathering together the terms
with the same order of $r$, we arrive at:

\begin{equation}  \label{eq36}
\sum\limits_k {\left[ {\left( {k + 1} \right)\left( {k + 2S + 1} \right)a_{k
+ 1} + \left( {\lambda - S - \frac{1}{2} - k} \right)a_k } \right]} r^k = 0.
\end{equation}

From (\ref{eq38}) we obtain the following recurrent relation for the
coefficients of (\ref{eq35}):

\begin{equation}  \label{eq37}
a_{k + 1} = \frac{\left( {k + S + \frac{1}{2} - \lambda } \right)}{\left( {k
+ 1} \right)\left( {k + 2S + 1} \right)}a_k .
\end{equation}

It is clear from this relation that at $r \to \infty $ the series (\ref{eq35}%
) diverges as $e^r$, and for its convergence the relation $\lambda = k + S +
\frac{1}{2}$ must be imposed. Using (\ref{eq26}), it can be presented as:

\begin{equation}  \label{eq38}
\frac{Z\alpha E}{\sqrt {\mu ^2c^4 - E^2} } = k + S + \frac{1}{2}.
\end{equation}

Solving the equation (\ref{eq38}) with respect to $E$, we obtain the energy
spectra of atom:

\begin{equation}  \label{eq39}
E_{k,m} = \frac{\left( {k + S + \frac{1}{2}} \right)\mu c^2}{\left[ {%
Z^2\alpha ^2 + \left( {k + S + \frac{1}{2}} \right)^2} \right]^{1 %
\mathord{\left/ {\vphantom {1 2}} \right. \kern-\nulldelimiterspace} 2}} =
\mu c^2\left[ {1 - \frac{Z^2\alpha ^2}{Z^2\alpha ^2 + \left( {k + \frac{1}{2}
+ \sqrt {m^2 - Z^2\alpha ^2} } \right)^2}} \right]^{1 \mathord{\left/
{\vphantom {1 2}} \right. \kern-\nulldelimiterspace} 2}.
\end{equation}

Finally, the solution of the (\ref{eq27}), which satisfies standard
conditions, can be expressed through the confluent hyper geometrical
function ${\ }_1F_1 \left( {\alpha ,\beta ;x} \right)$. Thus for the radial
waive function we can take

\begin{equation}  \label{eq40}
R_{n_\rho ,m} \left( r \right) = C_{n_\rho ,m} r^S\exp \left( {\ - r %
\mathord{\left/ {\vphantom {r 2}} \right. \kern-\nulldelimiterspace} 2}
\right){\ }_1F_1 \left( {S + \frac{1}{2} - \lambda ;\;2S + 1;\;r} \right),
\end{equation}

\noindent where $C_{n_\rho ,m} - $ is the normalization constant.

As it follows from (\ref{eq39}), the value of energy for $m = 0$ becomes
complex. Otherwise an instability, similar to one in three-dimensional
Klein-Gordon hydrogen atom at $Z\alpha > \frac{1}{2}$ [22-23], appears in
the system.

The relation (\ref{eq39}) could be obtained comparing of radial equations
for relativistic and non-relativistic problems and using the exact form of
energy spectra for non-realtivistic case. The equations are:
\begin{equation}  \label{eq41}
\begin{aligned} & - \frac{\hbar ^2}{2\mu }\frac{1}{\rho }\frac{d}{d\rho
}\left( {\rho \frac{d}{d\rho }} \right)R + \frac{\hbar ^2m^2}{2\mu \rho ^2}R
- \frac{Ze^2}{\rho }R = ER, \\ & - \frac{\hbar ^2}{2\mu }\frac{1}{\rho
}\frac{d}{d\rho }\left( {\rho \frac{d}{d\rho }} \right)R + \frac{\hbar
^2\left( {m^2 - Z^2\alpha ^2} \right)}{2\mu \rho ^2}R - \frac{Ze^2E}{\rho
\mu c^2}R = \frac{\left( {E^2 - \mu ^2c^4} \right)}{2\mu c^2}R. \end{aligned}
\end{equation}

It is easy to see that the second equation is reduced to the first one after
the following transformations:
\begin{equation*}
m^2 \to m^2 - Z^2\alpha ^2,\quad Z \to \frac{ZE}{\mu c^2},\quad E \to \frac{%
E^2 - \mu ^2c^4}{2\mu c^2}.
\end{equation*}
Substituting these notations into the expression for the energy of
non-relativistic two-dimensional hydrogen atom, we arrive at (\ref{eq39}).

As it was mentioned, Kane's dispersion law of the charge carriers in the
two-band approximation has the form analogous to the relativistic one. That
is why we can reduce the two-particle Hamiltonian function of the exciton
problem to the effective one-particle one, and apply the results obtained
above to the Kane's exciton.

For the Kane's exciton case the role of the light speed plays the parameter$s
$, mass of the free electron$\mu $ replaces with the effective mass $\mu _e $
of the electron in the crystal. From that for the effective constant of the
fine structure $\alpha ^ * $ we can take
\begin{equation}  \label{eq42}
\alpha ^ * = \frac{e^2}{\hbar s}.
\end{equation}

In the two-band interaction effective masses of the electron and hole are
equal,$\mu _e = \mu _h $, which makes it possible to bring the two-particle
Hamilton function to the one-particle. Indeed, passing to the new coordinate
system, in which $\mu _e \vec {\rho }_e + \mu _h \vec {\rho }_h = 0$, and
introducing $\vec {\rho } = \vec {\rho }_e - \vec {\rho }_h $, as well as
considering that $\vec {\rho }_e = \frac{\mu _h }{\mu _e + \mu _h }\vec {%
\rho }$ and $\vec {\rho }_h = - \frac{\mu _e }{\mu _e + \mu _h }\vec {\rho }$%
, for the electron and hole impulse operators we will get

\begin{equation}  \label{eq43}
\begin{aligned} \hat {\vec {P}}_e = \frac{\mu _e + \mu _h }{\mu _h^ * }\hat
{\vec {P}} \\ \hat {\vec {P}}_h = - \frac{\mu _e + \mu _h }{\mu _e }\hat
{\vec {P}}, \\ \end{aligned}
\end{equation}

\noindent where $\hat {\vec {P}} = - i\hbar \frac{\partial }{\partial \vec {%
\rho }}$. Switching to the new coordinate ${\vec {\rho }}^{\prime}= \frac{%
\vec {\rho }}{4}$ and introducing effective mass and charge, ${\mu }%
^{\prime}\equiv 2\mu ,\;{e}^{\prime}\equiv 2e$ we come to the following
equation

\begin{equation}  \label{eq44}
\left( \hat {P}^{\prime\; 2}s^2 + \mu^{\prime\; 2}s^4 \right)\Psi
_{n_\rho ,m} = \left( {E_{n_\rho ,m}^{ex} + \frac{{e}^{\prime\; 2}}{{\rho }%
^{\prime}}} \right)^2\Psi _{n_\rho ,m}
\end{equation}
\noindent which fully coincides with the Klein-Gordon equation for the
two-dimensional hydrogen-like atom, where instead of $\mu $ we have ${\mu }%
^{\prime}$, $e$ -- ${e}^{\prime}$, $c$ -- $s$, and ${\hat {\vec {P}}}%
^{\prime}= 2\hat {\vec {P}}$. This allows us to apply to the Kane's exciton
main results, obtained during the solution of the two-dimensional
relativistic hydrogen atom which are:

Exciton's energetic spectrum has the form

\begin{equation}  \label{eq45}
E_{n_\rho ,m}^{ex} = \left[ {1 - \frac{4\alpha ^{ * 2}}{4\alpha ^{ * 2} +
\left( {n_\rho + \frac{1}{2} + \sqrt {m^2 - 4\alpha ^{ * 2}} } \right)^2}} %
\right]{\mu }^{\prime}s^2,
\end{equation}

The states of the system with $m = 0$ are unstable.

Degeneration of energy levels, which takes place in the case of
two-dimensional exciton with the standard dispersion law, disappears.

Finally, after the standard procedure of the normalizations of the radial
part of the exciton's wavefunction
\begin{equation}  \label{eq46}
\int\limits_0^\infty {R_{n_\rho ,m}^2 \left( \rho \right)\rho d\rho = 1} ,
\end{equation}
and using the equation
\begin{equation}  \label{eq47}
\begin{aligned} I &=\int\limits_0^\infty {e^{ - \chi x}x^{\nu - 1}F^2\left(
{ - n;\gamma ;\chi x} \right)dx = \frac{\Gamma \left( \nu \right)n!}{\chi
^\nu \gamma \left( {\gamma + 1} \right) \cdots \left( {\gamma + n - 1}
\right)}} \\ &\times \left\{ {1 + \sum\limits_{p = 0}^{n - 1} {\frac{n\left(
{n - 1} \right) \cdots \left( {n - p} \right)\left( {\gamma - \nu - p - 1}
\right) \cdots \left( {\gamma - \nu + p} \right)}{\left[ {\left( {p + 1}
\right)!} \right]^2\gamma \left( {\gamma + 1} \right) \cdots \left( {\gamma
+ p} \right)}} } \right\}, \end{aligned}
\end{equation}
for $\Psi _{n_\rho ,m} \left( {\rho ,\varphi } \right) = R_{n_\rho ,m}
\left( \rho \right)\Phi _m \left( \varphi \right)$ we obtain:
\begin{equation}  \label{eq48}
\begin{aligned} \Psi _{n_\rho ,m} \left( {\rho ,\varphi } \right) &=
\frac{1}{a_{ex} }\sqrt {\frac{2\left( {2S + 1} \right)\mbox{...}\left( {2S +
n_\rho } \right)}{\pi \,\Gamma \left( {2S + 2} \right)n_\rho !\left( {1 +
\frac{2n_\rho }{2S + 1}} \right)\left( {Z^2\alpha ^2 + \left( {n_\rho + S +
\frac{1}{2}} \right)^2} \right)}} \\ &\times e^{im\varphi }e^{ - \varepsilon
\rho }\left( {2\varepsilon \rho } \right)^SF\left( { - n_\rho ;\,2S +
1;\,2\varepsilon \rho } \right). \end{aligned}
\end{equation}

\medskip

\begin{center}
\textbf{4. Conclusion }
\end{center}

So, the presence of hidden symmetry in two-dimensional Coulomb system
affects the character of forbidden exciton transitions in narrowband quantum
wells in the case when the system can be considered as an exact
two-dimensional one. The quasi-two-dimensionality or non-parabolisity of the
dispersion low takes away this hidden symmetry, and the explicit dependence
on magnetic quantum number $m$ appears. For two-band Klein model applied to
exciton levels (in this case the dispersion low of charge carriers is
similar to the relativistic one) apart from the removal of level degeneracy
with respect to $m$, instabilities for all energy levels since $m = 0$
appear in this problem.

\textbf{Acknoledgements. }Authors express their gratitude to Prof. L.G.
Mardoyan and Dr. T.S. Hakobyan for numerous and valuable discussions. This
work was supported by National Program of Armenian "Semiconductor Micro- and
Nanoelectronics" and  CRDF-NFSAT grant ARP1-3228-YE-04.

\begin{center}
\textbf{REFERENCES }
\end{center}

[1] \textit{D. Stehr,~M. Helm, C. Metzner, M. Wanke}. Phys. Rev. B. 2006. V.
74. id. 085311.

[2] \textit{C.~S. Liu,~H.~G. Luo, W.~C. Wu.} J. Phys.: Cond. Mat. 2006 .V.
18. P. 9659.

[3] \textit{J. Shumway.} Physica E. 2006. V. 32. P. 273-276.

[4] \textit{V. Savona,~ W. Langbein.}~ Phys. Rev. B. 2006. V. 74. id. 075311.

[5] \textit{Y. Feng,~X. Xu,~ H. Spector.} Physica E. 2006. V. 33. P. 201.

[6] \textit{V.~Vettchinkina,~ A. Blom,~K. Chao.} Phys. Rev. B. 2006. V. 72.
id. 045303.

[7] \textit{E.M. Kazaryan, A.A. Kostanyan, H.A. Sarkisyan.} Physica E. 2005.
V.28. P. 423.

[8] \textit{B.S. Monozon, P. Schmelcher.} Phys. Rev. B. 2005. V. 71. id.
085302.

[9] \textit{L.G.Mardoyan, G.S. Pogosyan, A.N. Sissakyan, V.M.Ter-Antonyan}.
Quantum systems with hidden symmetry. M.: Fizmatlit, 2006.

[10] \textit{B. Zaslow, M. Zandler.} Am. J. Phys. 1967. V. 35. P.1118.

[11] \textit{A. Cisneros, H. McIntosh.} 1969. J. Math. Phys. V. 10. P. 277.

[12] \textit{L.G.Mardoyan, G.S. Pogosyan, A.N. Sissakian, V.M.Ter-Antonyan}.
1984. Theor. and Math. Phys. V. 61. P. 99.

[13] \textit{A.I.Anselm.} Introduction to the theory of semiconductors. M.:
Science 1978.

[14] \textit{E.M. Kazaryan, R.L.Enfiajian.} 1971. Sov. Phys. Semicon. V. 5.
P. 2002.

[15] \textit{V.Halonen,~T. Chakraborty,~P. Pietil\'inen}.~Phys. Rev. B.
1992. V. 45. P.5980.

[16] \textit{K. Janssens, B. Ptroens, F. Peeters}. Phys. Rev. B. 2001. V.
64. P.155324.

[17] \textit{K. Janssens, B. Ptroens, F. Peeters}. Phys. Rev. B. 2002. V.
66. P.075314.

[18] \textit{M.Cidil'kovski.} Electrons and holes in semiconductors. M.:
Science 1972.

[19] \textit{G.M. Arutyunyan, M.G. Arutyunyan, G.S. Pogosyan,,
V.M.Ter-Antonyan.} Preprint of Laboratory of Theoretical Physics (N -77-10).
Yerevan, 1977.

[20] \textit{E.M. Kazaryan, A.P. Djotyan, H.A. Sarkisyan.} 1994 J. Cont.
Phys. V. 29, P. 90.

[21] \textit{L.D.Landau, E.M.Lifshic.} Quantun Mechanics. M.: Science, 1989.

[22] \textit{G. Bete.} Quantum Mechanics, Mir, 1983.

[23] \textit{V.B. Berestecki, E.M. Lifshic, L.P. Pitaevski. }Relativistic
quantum theory. M.: Science, 1971.

\end{document}